# Infrared-to-violet tunable optical activity in atomic films of GaSe, InSe, and their heterostructures

Daniel J. Terry,[1,2] Viktor Zólyomi,[1,2] Matthew Hamer,[1,2] Anastasia V. Tyurnina,[1,3] David G. Hopkinson,[4] Alexander M. Rakowski,[4] Samuel J. Magorrian,[1,2] Nick Clark,[4] Yuri M. Andreev,[5] Olga Kazakova,[6] Konstantin Novoselov,[1,2] Sarah J. Haigh,[1,4] Vladimir I. Fal'ko,[1,2] * Roman Gorbachev[1,2] **

[1] National Graphene Institute, Manchester, Oxford Road, M13 9PL, UK
[2] School of Physics and Astronomy, University of Manchester, Oxford Road, M13 9PL, UK
[3] Skolkovo Institute of Science and Technology, Nobel St. 3, 143026 Moscow, Russia
[4] School of Materials, University of Manchester, Oxford Road, M13 9PL, UK
[5] National Tomsk State Research University, 634050 Tomsk, Russia
[6] National Physical Laboratory, Teddington, TW11 0LW, UK

E-mail: *vladimir.falko@manchester.ac.uk   **roman@manchester.ac.uk

**Abstract**

Two-dimensional (2D) semiconductors - atomic layers of materials with covalent intra-layer bonding and weak (van der Waals or quadrupole) coupling between the layers[1] – are a new class of materials with great potential for optoelectronic applications[2]. Among those, a special position is now being taken by post-transition metal chalcogenides (PTMC), InSe and GaSe. It has recently been found[3] that the band gap in 2D crystals of InSe more than doubles in the monolayer compared to thick multilayer crystals, while the high mobility of conduction band electrons is promoted by their light in-plane mass. Here, we use Raman and PL measurements of encapsulated few layer samples, coupled with accurate atomic force and transmission electron microscope structural characterisation to reveal new optical properties of atomically thin GaSe preserved by hBN encapsulation. The band gaps we observe complement the spectral range provided by InSe films, so that optical activity of these two almost lattice-matched PTMC films and their heterostructures densely cover the spectrum of photons from violet to infrared. We demonstrate the realisation of the latter by the first observation of interlayer excitonic photoluminescence in few-layer InSe/GaSe heterostructures. The spatially indirect transition is direct in k-space and therefore is bright, while its energy can be tuned in a broad range by the number of layers.

Bulk gallium selenide has been studied for over 50 years[4,5]. It is a layered semiconductor with an optical band gap of $E_g \approx$ 2eV at room temperature (~2.1eV at 10K)[6,7]. Its structural layers consist of a mirror-plane-symmetric Se-Ga-Ga-Se arrangement shown in Fig. 1a, with adjacent layers separated by 0.798 nm in its most common ε-polytype[8]. The band structure of monolayer GaSe (with the unit cell structural motif $Ga_2Se_2$) has also been studied: theoretical predictions[9-11] using density functional theory (DFT), illustrated in Fig. 2a, include a conduction band with the edge at the Γ-point and a light electron mass, $m_{1L} \approx 0.17 m_e$, and a flat top valence band with a weak maximum off Γ-point, as confirmed recently by ARPES measurements[12].

The extension of DFT analysis onto thicker N-layer films of ε-GaSe, discussed in detail in SI (section S4), predicts similar features in the electronic spectrum of multilayers, as those seen in Fig. 2b and 2c, with even lighter conduction electrons and, most importantly, band gaps that strongly depend on the number of layers N. The band structures shown in Fig. 2 were obtained using VASP with the LDA pseudopotentials, followed by the 'scissor correction' to the value of the band gap (*i.e.*, adding 1 eV to the band gap for compensating for the difference between the LDA-predicted value and the gap measured in bulk crystals[8]). Up to now, experimental attempts to gauge such band gap variation in few-layer GaSe crystals have produced mixed results. No photoluminescence (PL) measurements have to date reported a significant change in the optical band gap compared to the bulk value[18-21]. However, scanning tunnelling spectroscopy has shown a significant widening of the electronic band gap from trilayer (2.3eV) to monolayer (3.5eV) for a molecular beam epitaxy grown GaSe/graphene heterostructure[12], and cathodoluminescence[22] measurements have given a broad peak at 3.3 eV for a CVD-grown GaSe monolayer. A deterioration of the structure of GaSe (e.g. due to oxidation in air) is the likely reason for the literature discrepancies, changing topography and Raman fingerprints within minutes of air contact[13-17]. Here, we report the first experimental observation of a pronounced dependence of the optical band gap on the thickness of the ε-GaSe film measured via micro-PL on pristine non-oxidised ε-GaSe films with thicknesses down to a single layer. The crystals we measure are protected by graphene or hBN encapsulation after exfoliation in an inert atmosphere and we find that the experimentally measured optical band gaps are in good agreement with the band structure calculations, as well as with the resonant behaviour of Raman scattering studied in the same samples. We also report the first observation of bright, room temperature interlayer excitonic photoluminescence in few-layer InSe/GaSe heterostructures with a peak energy tunable in a broad range by the number of layers.

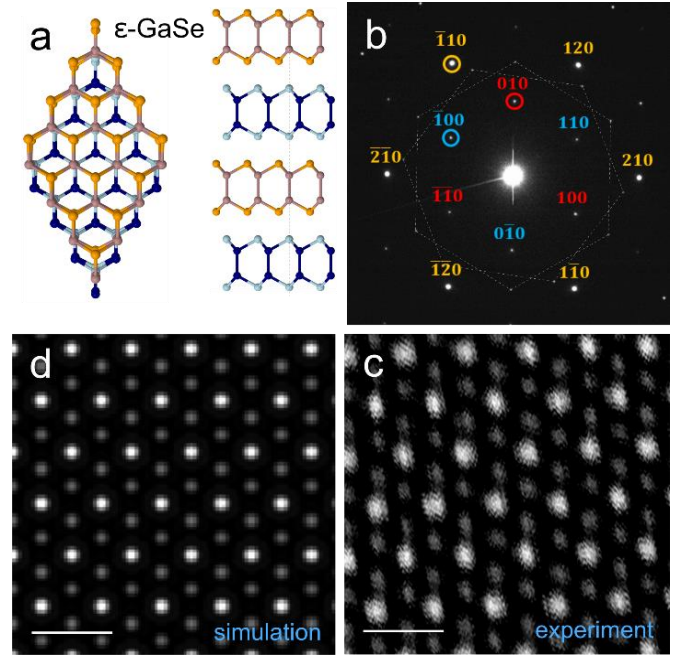

**Figure 1**: **(a)** Atomic structure and stacking arrangement of ε-GaSe. Orange and blue tones indicate equivalent stacking positions. Ga atoms shown as mauve and dark blue while Se atoms as orange and light blue. **(b)** Experimental electron diffraction pattern of few-layer GaSe. The circles correspond to the spot indices used to obtain the intensity ratios used for determining thickness and crystal polytype. The spots corresponding to the two encapsulating graphene sheets are indicated by hexagons. **(c)** Atomic resolution HAADF STEM images of few-layer GaSe encapsulated in graphene on both sides. **(d)** HAADF simulation for ε-GaSe polytype matching experimental observations. Scale bars 0.5 nm.

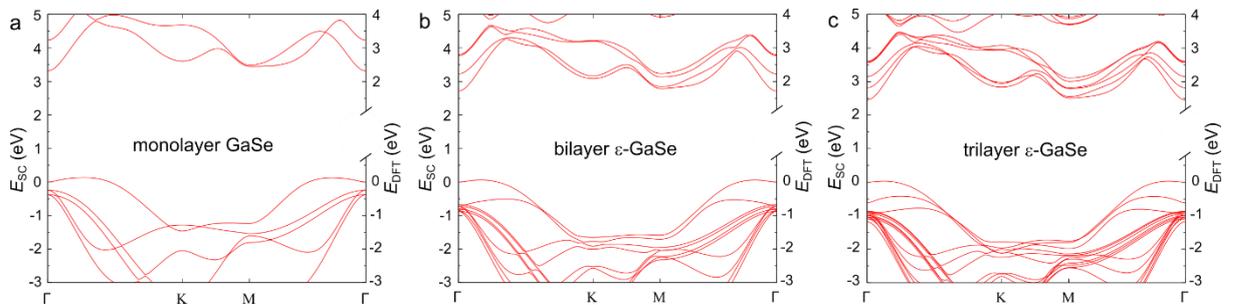

**Figure 2**: Band structure of monolayer GaSe and bi- and tri-layer ε-GaSe after a 'scissor correction' was applied (L.H.S. axis) to the spectra obtained using DFT with LDA pseudopotentials (R.H.S axis), based on the difference between the measured and calculated gap in bulk ε-GaSe.

Two-dimensional crystals studied in this work were exfoliated from a high-quality GaSe bulk crystal (nominally, ε-polytype) and fully encapsulated in either hexagonal boron nitride (hBN) or graphene in an inert environment[23].

and polarisation properties (parallel, ⇕→⇕ *versus* crossed, ⇕→⇔) prescribed by the crystal symmetry. Table I, which summarises experimental Raman data and DFT calculations for monolayer GaSe, shows that for all thicknesses (N=1,…6)

| IrrRep | Optical phonons in DFT | | Raman experiment | |
|---|---|---|---|---|
| | $\hbar\Omega_1$ [cm$^{-1}$] | Polarisation parallel : crossed | $\hbar\Omega_1$ [cm$^{-1}$] | $\hbar\Omega_{N\geq 2}$ [cm$^{-1}$] |
| E" | 55.4 | 0:0 | 0:0 | - | - |
| A$_1$' | 129.8 | 1:0 | 1:0 | 134.2 | 135.7 |
| E" | 211.6 | 0:0 | 0:0 | - | - |
| E' | 217.0 | 1:1 | 1:1 | 214.5 | 214.5 |
| A$_2$" | 248.3 | 0:0 | 0:0 | - | - |
| A$_1$' | 312.5 | 1:0 | 1:0 | 310.3 | 308.8 |

**Table 1**: Raman activity of phonons in atomically thin films of ε-GaSe. Phonon are classified according to the irreducible representation (IrrRep) of the symmetry group D$_{3h}$ of the monolayer crystal. Theoretical values of phonon energies $\hbar\Omega_1$ in the monolayer and polarisation selection rules for Raman scattering [parallel *versus* (:) crossed] are shown on the left (1 stands for 'active', 0 – 'inactive', 1:1 means 'equal strength'). Experimental results on Raman, taken from Fig. 3a, are on the right, showing a 1.5 cm$^{-1}$ shift between the multi-layer and monolayer A$_1$' modes.

Altogether, 17 samples of encapsulated thin GaSe films were fabricated and studied, with 100 individually identified regions of different thicknesses. As the bulk form of GaSe exists in four different polytypes (β, ε, δ and γ)[4,8,24], we carried out transmission electron microscopy imaging and electron diffraction investigations (Fig. 1) to establish the exact crystal structure of the material used. For this, we transferred graphene-encapsulated GaSe samples onto a silicon nitride support grid and obtained high angle annular dark field (HAADF) scanning transmission electron microscope (STEM) images of several few-layer GaSe specimens (a typical image is shown in Fig. 1c). The observed contrast is consistent with either ε-GaSe or δ-GaSe (see Fig. 1d for the simulated HAADF STEM image for the ε-polytype and SI for δ, β and γ-polytypes), while further analysis of the intensities of the diffraction spots allows us to rule out the δ-GaSe structure and to identify the polytype as ε-GaSe as well as to confirm the precise number of layers and high crystal quality (see SI for details).

AFM characterisation of encapsulated GaSe samples show atomically flat terraces with a step height changing in integer multiples of the bulk interlayer separation 0.8 nm. In contrast, exposed regions of GaSe are typically thicker, up to an additional 2-3 nm even for single layer crystals (see SI Figure 11). The integrity of the encapsulated GaSe terraces was further confirmed by both STEM and micro-Raman characterisation, see Fig. 3a, showing that even the monolayer regions of GaSe films maintain their integrity. The observed Raman signal is consistent with the Raman spectra taken on bulk crystals[14,17,19], with the Γ-point optical phonon energies,

Raman spectra contain two A'-type modes which are active only in parallel (incoming-scattered light) polarisations, and one E' type mode present in both parallel and crossed polarisations measurements. For the non-degenerate A' phonons (z→-z symmetric vibrations), described by a 'scalar' amplitude, $w$, the Raman interaction is governed by a nonlinear coupling $\gamma_A|\mathcal{E}|^2 w$ which prescribes the preservation of photon's polarisation in its inelastic scattering. For the double-degenerate E' phonon, described by a z→-z symmetric sublattice displacement vector, $(u_x, u_y)$, the Raman interaction of in-plane polarised photons is determined by coupling in the form of $\mathcal{E}^2 u$, where $\mathcal{E} = \mathcal{E}_x + i\mathcal{E}_x$ and $u = u_x + iu_x$ (in a crystal with symmetry D$_{3h}$, the 3$^{rd}$ order rotational symmetry axis makes angular momenta $3\hbar$ and 0 identical); this determines equal intensities for inelastic light scattering with parallel and crossed polarisations. At the same time, z→-z asymmetric modes A" and E" would require the involvement of the out-of-plane polarisation component of the photon field, which was impossible in the geometry of the Raman set-up used in the reported experiments.

Figure 3b shows PL spectra recorded on individual GaSe terraces of various thicknesses (here, PL intensity is normalised by the number of layers, to account for a higher absorption by thicker crystals, and substrate contributions are removed, see SI for details). For GaSe films with N>15, the PL peak is at 1.992 ± 0.006eV, as in bulk GaSe[20,21,25]. Similar to InSe films[3], thinner layers of GaSe show a strong blue shift of the PL peak, measured down to a bilayer crystal, as illustrated in Fig. 3c. Despite a very clear Raman signal in monolayer films, we were not able to detect PL in monolayer crystals up to an excitation of 3.8eV. We attribute this lack of PL activity to the shift of the valence band maximum away from the Γ-point, see Fig. 2, causing a shift towards an indirect transition for the monolayer. Additionally, the mirror symmetry of the monolayer crystal suppresses coupling of in-plane polarised photons with the interband optical transition (optical activity is possible only due to interband spin-split transitions facilitated by spin-orbit coupling). To note, the experimentally measured PL spectra correlate well with the resonance behaviour of the Raman scattering in the same films, such that an enhancement in the Raman intensity is observed when the laser excitation approaches the PL energy for a specific crystal thickness (Fig. 3d).

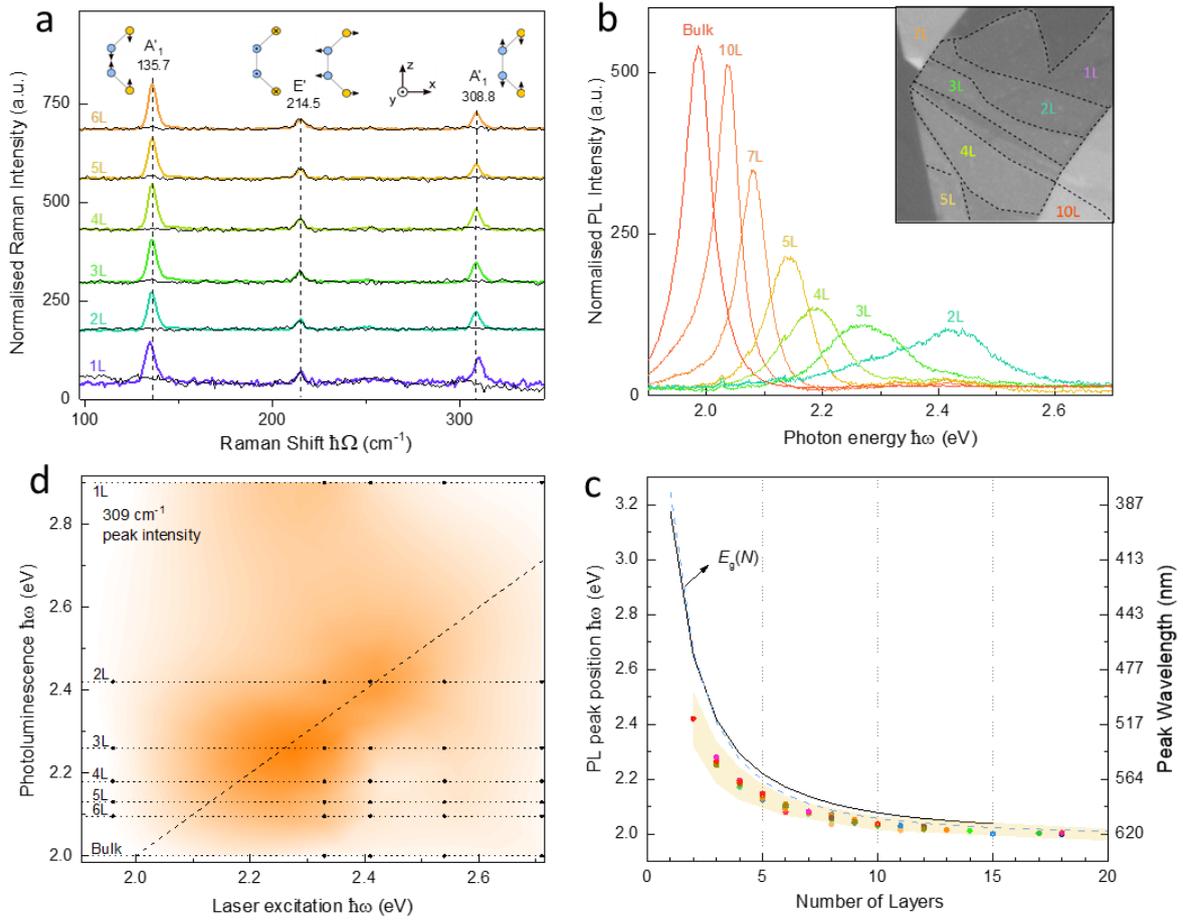

**Figure 3**: **(a)** Raman spectra for various GaSe thicknesses obtained with a 2.33eV pumping laser. Coloured spectra correspond to the parallel polarisation condition; black spectra correspond to cross-polarisation condition. Illustrations show corresponding displacements for Ga (blue) and Se (yellow) atoms. **(b)** PL spectra for different thicknesses obtained on a single sample with multiple terraces, normalised by the corresponding number of layers measured with a 3.06 eV laser. Inset shows optical image of the encapsulated sample (image width 20 µm). **(c)** PL peak energy as a function of the number of layers obtained experimentally from 70 regions of various thicknesses (coloured circles represent different samples), compared to the $E_g$ calculated using scissor corrected DFT (black line) and hybrid k.p theory tight-binding model (blue line). Orange shading indicates experimental linewidth at half maximum. **(d)** Intensity of 309 cm$^{-1}$ Raman peak as a function of laser energy and PL peak position for each number of layers (white to orange from 0 to 30 counts). Dashed line indicates condition when laser excitation is equal to the PL peak position.

The observed dependence of the PL peak position on the number of layers in the 2D crystal, down to N=4,3,2, is shown in Fig. 3c by coloured symbols in comparison with the theoretically calculated values of the band gaps in ε-GaSe films of various thicknesses. There are two theoretical predictions for $E_g$ displayed as solid lines Fig. 3c. The black line corresponds to the band gaps taken from the multilayer bands computed using VASP, such as given in Fig. 2. The second set (blue dashed line) was obtained from a hybrid k.p theory tight-binding model replicating similar calculations performed for monolayers and multilayers of InSe[26,27]. In the former approach we have implemented a 'scissor correction', shifting conduction and valence bands to match $E_g$ known for bulk GaSe. In the latter approach, discussed in detail in SI, we parameterised the k.p theory expansion of the intra-layer two-band Hamiltonians (including the value of the monolayer gap) and interlayer hopping was parameterised to reproduce the DFT dispersion along Γ-A dispersion in the bulk crystal and conduction/valence band dispersions near Γ-point in the monolayer, and the experimentally known band gap in the bulk material. These theoretically calculated values are slightly higher than the experimentally measured PL peaks positions: we attribute that difference (0.05-0.3eV) to the exciton binding energies, which are larger for thinner films due to both weaker dielectric screening and stronger quantum confinement in a thinner material.

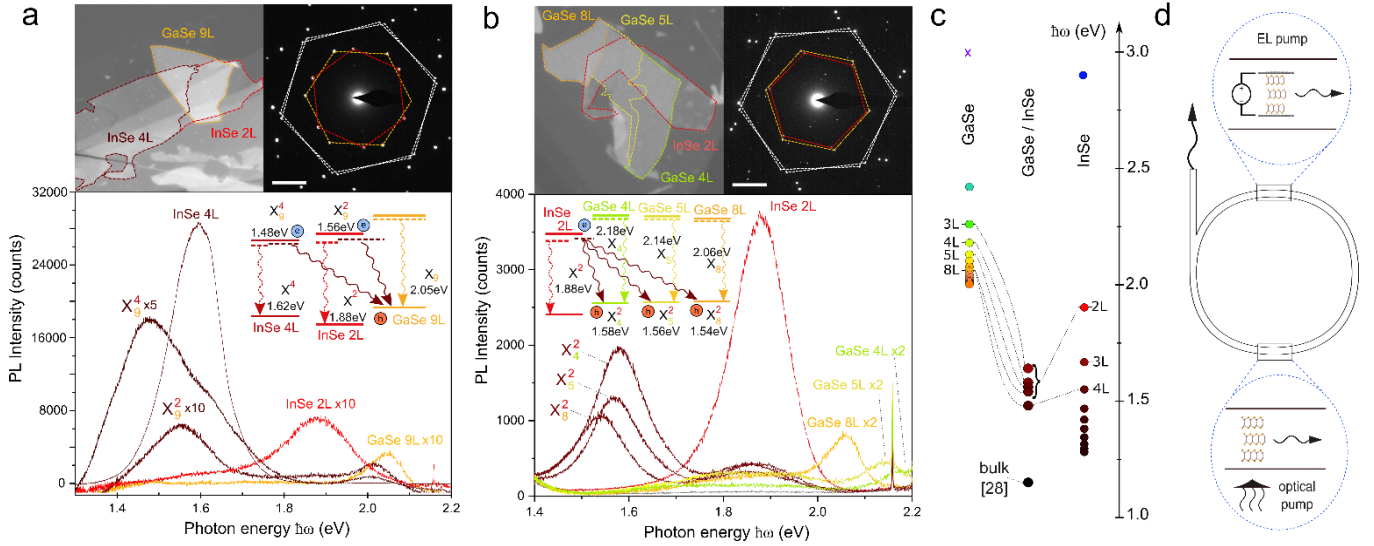

**Figure 4:** (**a, b**) Top: optical images of encapsulated InSe/GaSe heterostructures, outlined are the InSe bilayer film (red) and GaSe terraced film (orange-yellow), image width 30 µm (**a**) and 25 µm (**b**), with corresponding selected area electron diffraction patterns demonstrating the twist angles between the PTMCs (InSe red, GaSe orange, hBN white); scale bar 2nm$^{-1}$. Bottom: PL spectra for encapsulated N-layer GaSe and M-layer InSe films as well as their respective overlapping heterostructure regions showing interlayer excitonic emission $X_{M\ GaSe}^{N\ InSe}$, note that the peak at ~2.16 eV is due to a hBN Raman mode (for Raman data see SI Figure 14). The indirect transition in the heterostructures is blue-shifted from the 1.1 eV line observed earlier in a van der Waals interface between bulk InSe and bulk GaSe[28]. Inset: Illustration shows corresponding band alignment of the n-InSe and GaSe heterostructure giving the interlayer exciton energy. Solid lines are band edges, dashed lines are excitonic states (**c**) Summary of measured samples showing the PL peak energy. (**d**) Concept schematic of light emitting devices based on InSe, GaSe and their heterostructures. Emission can be produced by either optical pumping or an external voltage, and can also be placed in an optical cavity to create a laser.

In the following, we discuss encapsulated few-layer InSe/GaSe heterostructures. Figure 4a shows optical images, selected area electron diffraction (SAED) patterns and PL spectra obtained at room temperature on an encapsulated 9-layer GaSe partially overlapping with terraced 2 and 4-layer InSe crystal. Apart from the expected PL emission from each individual material away from the overlap, we observe a new luminescence peak from the overlap region. The energies of these transitions, denoted as $X_9^2$ and $X_9^4$ (indexes stand for number of layer in each material, $X_{M\ GaSe}^{N\ InSe}$), are lower than those of its constituent PTMC films individually but higher than the indirect exciton at 1.1 eV observed at an interface of the bulk materials[28]. The InSe has residual n-doping and GaSe is undoped, therefore a type-II band alignment occurs allowing for spatially indirect excitonic emission – as observed previously in heterobilayers of TMDCs[29,30] (and now in another set of 2D materials). Here we also observe that the energy of the indirect excitons is tuned by choosing different combinations of PTMC layer thicknesses of either of the two 2D materials in the heterostructures. Thus, in the Fig. 4a the thickness of InSe varied from 2 to 4 to observe a notable shift in the intralayer exciton energy ~80 meV between $X_9^2$ and $X_9^4$, while in Fig. 4b we demonstrate that the thickness of GaSe changes to reveal gradual tuning between $X_4^2$, $X_5^2$ and $X_8^2$. Comparison of the interlayer transition with those of individual crystals allows us to plot approximate optical band alignment as shown in the inset of Fig. 4a,b. The intensities of the interlayer excitonic photoluminescence are also found to be between the brightness of their constituent PTMC layers, irrespective of alignment. Here, the relative alignment of two heterostructures of similar PTMC film thicknesses ($X_9^2$ and $X_8^2$) were measured via SAED, revealing twist angles of ~21.7° and ~2.7°; see Fig. 4a and 4b respectively. Although a significant difference in the twist angle is measured, a shift of only ~20meV is observed between the interlayer exciton energies (within the range of variation of energies seen in individual films of the same thickness; see Fig. 3c) as well as the interlayer excitonic emission remaining bright. We suggest that this is due to the close proximity of the band extrema in both PTMCs to the Γ-point; see Fig. 2 and ref. 10, 26. Thus, the relative alignment of the PTMC layers is expected to have a negligible effect upon the recombination in k-space, allowing the creation of bright interlayer excitons at room temperature without the need for careful angle alignment and lattice matching[31,32]. While the interlayer exciton dominates spectra in the overlap region, a reduced intensity of individual PL peaks can be often seen, probably due to the presence of a small amount of contamination in pockets, size <1µm, between the GaSe and InSe crystals, possibly resulting in a reduced interlayer exciton luminescence intensity as well.

The observed evolution of optical spectra of thin films of GaSe demonstrates a strong dependence of the band gap on the number of layers in these 2D crystals. By combining the data for the photons emitted in the course of interband transitions in thin films of GaSe and InSe, Fig. 4c, we find that the range of room-temperature optical activity of thin films of these two materials densely covers the interval from $E_g$ ~1.3eV to 2.5eV. This interval can be further extended down to 1.1 eV by using N layer GaSe / M layer InSe type-II heterostructures[28], and the density of spectral coverage could be increased by using thin films of $Ga_xIn_{2-x}Se_2$ alloys. As a result, thin films of PTMCs provide a versatile materials platform for applications in photodetectors[20], second harmonic generation[33], and single-photon emission[34]. Moreover, the analysis of the DFT-calculated wave functions at the band edges of $Ga_{2N}Se_{2N}$ (see in SI) and $In_{2N}Se_{2N}$[26] films shows that the interband transitions in these systems are strongly coupled to photons polarised perpendicular to the 2D crystal. This suggests that if such layers were implemented in optical fibre waveguides over the areas with the highest electric field component of the photon field, they could be used as the excitation source in ring-lasers. As sketched in Fig. 4d, PTMC films can be excited either by optical pumping using higher-energy photons[3], or, even more interestingly, by the electrical injection of electrons and holes in vertical tunnel diode structures G/hBN/PTMC/hBN/G or G/InSe/GaSe/G using few-layer graphene (G) as the electrodes (which would be optically passive for the proposed polarisation arrangement of the optical circuit) and hBN as the injection barriers.

## Methods

In order to prevent oxidation, our samples were fabricated using a specially designed micromanipulation setup placed inside an argon filled glovebox with <0.1ppm $O_2$ and $H_2O$. First, bulk PTMC crystals were mechanically exfoliated onto a 200nm layer of poly-(propylene carbonate) (PPC) which had been previously spin coated onto a Si wafer. Optical microscopy was used for thickness identification and selected PTMC crystals were then picked up with hBN films residing on a polymer membrane using the dry peel transfer technique[23]. The PTMC/hBN heterostructures were then transferred onto another hBN film exfoliated onto an oxidised silicon wafer to achieve the full encapsulation. These heterostructures were finally removed from the glovebox to perform PTMC thickness verification via atomic force microscopy (AFM). AFM mapping was conducted routinely on each sample to confirm homogeneity and select regions free of structural defects such as cracks and folds. Alternatively, graphene was used for encapsulation of TEM samples due to its availability and larger lateral size.

All PL measurements for films of >2L were taken at room temperature using either a Horiba XploRA PLUS system with a continuous wave (CW) 532nm laser or a Horiba LabRAM HR Evolution system with CW 488nm laser. For PL on both of these Horiba systems a spot size of <1µm, grating of 600 groves per millimetre, 100x lens (NA = 0.9) and exposure time of 1s was used. A maximum incident laser power of ~4.8mW was used for the PL measurements on the thinnest samples. For the 2L film (see Fig. 3a) a custom-built system using a CW 405nm laser, spot size <1µm, grating of 300 groves per millimetre, 100x lens (NA = 0.9) and exposure time of 3s was used. Here, a maximum incident laser power of ~1.1mW was applied with the PL averaged over 3 accumulations. The use of the 488nm and 405nm laser enabled the PL detected from the 3L and 2L films to be isolated from the Raman spectra. For the thinner films (<5L), PL maps were also produced in order to confirm that the reduced PL intensities were localised to corresponding layers and not an outlying measurement (see Supplementary Section 7 for demonstrations of this). PL measurements from monolayers of GaSe were also attempted using a CW 325nm laser on the Horiba LabRAM HR Evolution system, with a spot size ~2µm, a power of ~1.2mW, 40x lens (NA = 0.5) and 600 groves per millimetre.

Raman spectra were taken for each sample in order to confirm the presence of GaSe as well as a lack of oxidation. The Raman data was collected using the same spectrometers with laser excitation energies of 1.96, 2.33, 2.41, 2.54 and 2.7eV, grating of 2400 groves per millimetre and the objective of 100x magnification. Depending on the Raman setup and sample thickness, the laser power and acquisitions setting were adjusted in order to obtain reasonable signal to noise ratio. Additionally, the spectrum from an area near each film was subtracted from all GaSe Raman spectra due to the presence of background from wide Si band around 308cm$^{-1}$. The intensities in Fig. 3a were normalised by intensity of the E' mode at 215cm$^{-1}$. All intensities from the map plot on the Fig. 3d were then normalised by the number of layers, laser excitation power and time of spectral acquisition.

**Author contributions**

V.I.F., K.S.N., O.K. and R.V.G. conceived the study; D.J.T. fabricated devices with help from M.H.; S.J.H., A.M.R. and D.G.H. performed electron imaging and diffraction; D.J.T. and A.V.T. performed optical measurements and analysed experimental data; V.Z., S.J.M. and V.I.F. provided theory used for the interpretation of the experiments; Y.M.A. and O.K. provided bulk GaSe crystals, V.I.F. and R.V.G. wrote the manuscript with the help of D.J.T., A.V.T. and V.Z.; S.J.M, V.Z, D.J.T. and M.H. wrote the Supporting Information, and all authors contributed to discussions.

**Acknowledgments:**

We thank M. Danovich, M. Potemski, F. Koppens, M. Molas, S. Heeg, A. Patane, E. Tóvári and J. Howarth for useful discussions. This study was supported by the European Graphene Flagship Project; EPSRC grant EP/N010345; ERC Synergy Grant Hetero2D; EPSRC CDT Graphene-NOWNANO EP/L01548X, and fabrication of devices was performed using facilities of Henry Royce Institute for Advanced Materials. R.V.G. acknowledges financial support from the Royal Society Fellowship Scheme. S.J.H. acknowledges financial support from the ERC Starter Grant EvoluTEM, EPSRC grants EP/P009050/1 and EP/M010619/1 and from the US Defence Threat Reduction Agency (HDTRA1-12-1-0013). D.G.H., D.J.T., M.H., S.J.M. and A.M.R. acknowledge support from the EPSRC NowNano Doctoral Training Centre. D.J.T. also acknowledges financial support from Samsung Advanced Institute of Technology (SAIT). V. Z. and V. I. F. acknowledge support from the N8 Polaris service and the use of the ARCHER supercomputer (RAP Project e547).